\documentclass{article}
\usepackage{amssymb,amsmath,epsfig,epstopdf,color,graphicx,subfigure}
\usepackage{epstopdf,array}
\newcolumntype{L}{>{\centering\arraybackslash}m{3cm}}

\usepackage{arxiv}

\usepackage[utf8]{inputenc} % allow utf-8 input
\usepackage[T1]{fontenc}    % use 8-bit T1 fonts
\usepackage{hyperref}       % hyperlinks
\usepackage{url}            % simple URL typesetting
\usepackage{booktabs}       % professional-quality tables
\usepackage{amsfonts}       % blackboard math symbols
\usepackage{nicefrac}       % compact symbols for 1/2, etc.
\usepackage{microtype}      % microtypography

\usepackage{amsmath,amssymb,amsfonts}
\usepackage{algorithmic}
\usepackage{graphicx}
\usepackage{textcomp}
\usepackage{comment,color}
\usepackage{enumitem}
\usepackage{epsfig}
\usepackage{caption}
\usepackage{verbatim,bm}
\usepackage{epstopdf,array,mathtools,url,slashbox,subfigure}

\newcolumntype{L}{>{\centering\arraybackslash}m{3cm}}
\DeclarePairedDelimiter\norm{\lVert}{\rVert}%
\makeatletter
\let\oldabs\abs
\def\abs{\@ifstar{\oldabs}{\oldabs*}}
\let\oldnorm\norm
\def\norm{\@ifstar{\oldnorm}{\oldnorm*}}
\makeatother
\title{Solving the acoustic VTI wave equation using physics-informed neural networks}

\author{
  Chao Song\\
 Physical Sciences and Engineering Division\\
 King Abdullah University of Science and Technology
    \And
 Tariq Alkhalifah\\
 Physical Sciences and Engineering Division\\
 King Abdullah University of Science and Technology
  \And
  Umair bin Waheed\\
  Department of Geosciences\\
  King Fahd University of Petroleum and Minerals
}

\begin{document}
\maketitle

\begin{abstract}
Frequency-domain wavefield solutions corresponding to the anisotropic acoustic wave equations can be used to describe the anisotropic nature of the earth. To solve a frequency-domain wave equation, we often need to invert the impedance matrix. This results in a dramatic increase in computational cost as the model size increases. It is even a bigger challenge for anisotropic media, where the impedance matrix is far more complex. To address this issue, we use the emerging paradigm of physics-informed neural networks (PINNs) to obtain wavefield solutions for an acoustic wave equation for transversely isotropic (TI) media with a vertical axis of symmetry (VTI). PINNs utilize the concept of automatic differentiation to calculate its partial derivatives. Thus, we use the wave equation as a loss function to train a neural network to provide functional solutions to form of the acoustic VTI wave equation. Instead of predicting the pressure wavefields directly, we solve for the scattered pressure wavefields to avoid dealing with the point source singularity. We use the spatial coordinates as input data to the network, which outputs the real and imaginary parts of the scattered wavefields and auxiliary function. After training a deep neural network (NN), we can evaluate the wavefield at any point in space instantly using this trained NN. We demonstrate these features on a simple anomaly model and a layered model. Additional tests on a modified 3D Overthrust model and a model with irregular topography also show the effectiveness of the proposed method. 

\end{abstract}

% keywords can be removed
\keywords{Frequency domain, acoustic wave equation modeling, VTI media, physics-informed neural network.}

\section{Introduction}

Frequency-domain wave equation modeling is an important topic in seismic exploration. By inverting a potentially large impedance matrix of the wave equation, which is also referred to as the Helmholtz equation, the resulting Green's function can be used to illuminate the structures of the subsurface using many techniques, like reverse time migration (RTM) or full-waveform inversion (FWI). A simple acoustic isotropic assumption of the Earth often causes poor results in many areas that has strong anisotropy. \cite{alkhalifah2000acoustic} derived an acoustic wave equation in transversely isotropic media with a vertical symmetry axis (VTI media) using an acoustic dispersion relation obtained by setting the vertical shear wave velocity to zero \cite{alkhalifah1998acoustic}. To simplify the original fourth-order differential equation, \cite{zhou2006anisotropic} proposed a set of second-order wave equations for VTI media. These new acoustic VTI wave equations introduce auxiliary functions to form two second-order differential equations, and these two equations are easier to solve than the original fourth order formula \cite{song2020efficient}. 

However, solving this new acoustic VTI wave equation in the frequency domain requires inverting a large impedance matrix to include the pressure wavefield and auxiliary parameter. This will result in a tremendous burden on our computing resources, and makes it challenging to obtain wavefield solutions for large models, especially for 3D cases. The acoustic wave equation for anisotropic media suffers severely from the shear wave artifacts, especially for cases in which the sources are located in the anisotropic region \cite{alkhalifah2000acoustic}. These artifacts can be reduced by placing the sources in the isotropic layer, like in the case of marine survies. However, these artifacts are inevitable for some applications, such as microseismic event estimation \cite{shi2018microseismic} and reflection-waveform inversion (RWI) \cite{wu2016waveform}, when the sources (or secondary sources) are in the subsurface. In addition, there is another challenge for the commonly used finite-difference methods. It is difficult to deal well with models having irregular topography. Thus, it is important to seek an alternative way to obtain wavefield solutions, especially for anisotropic media. 

Machine learning (ML) is quickly gaining a lot of attention due to its ability to deal with large data. The support vector machine (SVM) is an effective approach widely used in pattern recognition and classification \cite{vapnik2013nature}, and it has been applied to AVO classification \cite{li2004support}, seismic facies recognition \cite{zhao2015comparison}, source type classification \cite{song2018source}, and image artifacts suppressing \cite{microsvm,chen2020suppressing}. With the rapid developments in computing capabilities and the rampant growth of available data, neural networks (NN) and different forms of it (e.g., deep/convolutional/recurrent NN) are gaining more attention. By taking advantage of abilities of the convolutional NN in image processing, it has shown effectiveness in detecting salt bodies \cite{shi2018automatic}, horizons, and faults \cite{wu2019faultseg3d}. Besides classification applications, convolutional NN can be used to predict low-frequency data to better enable FWI to converge \cite{ovcharenko2019deep} and monitor the time-lapse velocity change \cite{regone2017geologic}. Deep NN has also been utilized to pick seismic arrival time \cite{zhu2019phasenet}, improve the resolution of the migrated images \cite{kaur2020improving} and FWI inverted velocity models \cite{zhang2020high,liyy}. Seismic wave simulations \cite{siahkoohi2019TRnna,moseley2020solving} and direct velocity inversions can be solved by deep NN \cite{8931232,ren2020physics}. Most of the previous work rely on building the connections between the input and output data. This kind of supervised learning often requires a large amount of data and the corresponding manual labelling in the training process. It is also very dependent on the training set. 

Recently, a framework referred to as physics-informed neural networks (PINNs) has been proposed to solve partial differential equations (PDEs) \cite{raissi2019physics}. PINNs use physical laws in the loss function instead of pure data-mapping objectives. Using the concept of automatic differentiation, we are able to calculate the partial derivatives with respect to  spatial and temporal coordinates within the networks. PINNs demonstrated successful applications in the cardiac activation mapping \cite{sahli2020physics} and qualitative flow fields characterizing \cite{raissi2020hidden}. In geophysics, researchers have successfully applied PINNs in solving the time- and frequency-domain wave equations \cite{moseley2020solving,helmholtz} and isotropic P-wave eikonal equation \cite{eikonal}. In solving the Helmholtz equation using PINNs, they propose to solve for the scattered wavefield instead of solving the wave equation directly to avoid the point source singularity. The isotropic acoustic wavefields cannot accurately describe the wave propagation due to the anisotropic nature of the Earth. Solving an anisotropic wave equation requires additional computational cost. Analogous to previous work on solving the Helmholtz equation using PINNs, we can further use PINNs to predict the pressure wavefields for acoustic VTI media by using an acoustic VTI wave equation as a loss function.

In this paper, we propose a novel method to solve the scattered wavefield corresponding to the acoustic VTI wave equation using PINNs. The background wavefield solutions used in the scattered acoustic VTI wave equation can be obtained by analytical solutions with little computational cost. We use spatial coordinate values as input data to the network, and consider the velocity and anisotropic parameters as the complementary variables in the loss function. We build a fully connected deep neural network to train the network. After training the network, we can predict the real and imaginary parts of the scattered wavefields for each location in space. Applications on an anomaly as well as a layered model show that the proposed method is able to generate acoustic wavefield solutions free of shear wave artifacts. An application on a 3D Overthrust model also demonstrates the effectiveness of the proposed method.

\section{Theory}

\subsection{The acoustic VTI wave equation}

Considering the anisotropic nature of the earth, anisotropic acoustic wave equations are used to simulate wave propagation that approximately represents (at least kinematically) the behavior of P-waves inside the Earth. In acoustic VTI media, we can solve a coupled system of second-order differential equations to get the frequency-domain wavefields. The 3D frequency-domain acoustic VTI wave equation with constant density parameterized using the normal moveout (NMO) velocity $v_{n}$ and the anisotropic parameters $\delta$ and $\eta$, is stated as follows \cite{zhou2006anisotropic}:
\begin{eqnarray}
&&\omega ^{2}m_{n}p+\frac{\partial ^{2}(p+q)}{\partial x^{2}} +\frac{\partial ^{2}(p+q)}{\partial y^{2}}+\frac{1}{(1+2\delta)}\frac{\partial ^{2}p}{\partial z^{2}}=s,\nonumber\\
&&\omega ^{2}m_{n}q+2\eta \frac{\partial ^{2}(p+q)}{\partial x^{2}}- \frac{\partial ^{2}(p+q)}{\partial y^{2}}=0,
\label{eqn:eq1}
\end{eqnarray}
where $p$ is the pressure wavefield, and $q$ is an auxiliary perturbation wavefields. $s$ denotes the source function. $\omega$ represents the angular frequency. We use $m_{n}=\frac{1}{v_{n}^{2}}$ to represent the squared slowness. As we mentioned in the introduction section, our objective is to solve for the scattered pressure wavefield $\delta p$, which is defined as:
\begin{eqnarray}
\delta p=p-p_{0},
\label{eqn:eq2}
\end{eqnarray}
where $p_{0}$ is the background wavefield satisfying the isotropic wave equation, which is given by,
\begin{eqnarray}
&& \omega ^{2}m_{n0}p_{0}+\frac{\partial ^{2}p_{0}}{\partial x^{2}}+\frac{\partial ^{2}p_{0}}{\partial y^{2}}+\frac{\partial ^{2}p_{0}}{\partial z^{2}}=s,
\label{eqn:eq3}
\end{eqnarray}
where $m_{n0}=\frac{1}{v_{n0}^{2}}$ represents the squared slowness corresponding to the background model, which we set to be homogeneous for simplicity. The background anisotropic parameters $\eta_{0}$ and $\delta_{0}$ are zero. In an isotropic acoustic medium, the auxiliary function $q$ equals zero. The 3D isotropic acoustic wave equation in equation \ref{eqn:eq3} allows an analytical solution for a constant velocity and a point source located at $\mathbf{x_{s}}$, which is given by:
\begin{eqnarray}
p_{0}(\mathbf{x})=\frac{e^{i\omega \sqrt{m_{n0}}\left | \mathbf{x-x_{s}} \right |}}{\left | \mathbf{x-x_{s}} \right |},
\label{eqn:eq4}
\end{eqnarray}
where $\mathbf{x}=\left \{ x,y,z \right \}$ defines the spatial coordinates in the Euclidean space. For the 2D case, the analytical solution for the isotropic acoustic wave equation is expressed as:
\begin{eqnarray}
p_{0}(\mathbf{x})={iH_{0}^{(1)}(\omega \sqrt{m_{n0}}\left | \mathbf{x-x_{s}} \right |}),
\label{eqn:eq5}
\end{eqnarray}
where $H_{0}^{(1)}$ is the Hankel function of the first kind and order 0 \cite{engquist2018approximate}. If we insert $p=p_{0}+\delta p$ into equation \ref{eqn:eq1}, we obtain a relation between $p_{0}$, $\delta p$ and $q$, as shown:
\begin{eqnarray}
&& \omega ^{2}m_{n}(p_{0}+\delta p)+\frac{\partial ^{2}(p_{0}+\delta p+q)}{\partial x^{2}} +\frac{\partial ^{2}(p_{0}+\delta p+q)}{\partial y^{2}}\nonumber\\
&& + \frac{1}{1+2\delta }\frac{\partial ^{2}(p_{0}+\delta p)}{\partial z^{2}}=s,\nonumber\\
&& \omega ^{2}m_{n}q+2\eta \frac{\partial ^{2}(p_{0}+\delta p+q)}{\partial x^{2}}=\frac{\partial ^{2}(p_{0}+\delta p+q)}{\partial y^{2}}.
\label{eqn:eq6}
\end{eqnarray}
We define the squared slowness perturbation as $\delta m_{n}=\frac{1}{v_{n}^{2}}-\frac{1}{v_{n0}^{2}}$. We subtract equation \ref{eqn:eq3} from equation \ref{eqn:eq6}, and the scattered wavefield $\delta p$ in acoustic VTI media then satisfies
\begin{eqnarray}
&& \omega ^{2}m_{n}\delta p+\frac{\partial ^{2}(\delta p+q)}{\partial x^{2}}+\frac{\partial ^{2}(\delta p+q)}{\partial y^{2}}+\frac{1}{1+2\delta }\frac{\partial ^{2}\delta p}{\partial z^{2}}= \nonumber\\
&& -\omega ^{2}\delta m_{n} p_{0}-(\frac{1}{1+2\delta }-1)\frac{\partial ^{2}p_{0}}{\partial z^{2}},\nonumber\\
&& \omega ^{2}m_{n}q+2\eta \frac{\partial ^{2}(\delta p+q)}{\partial x^{2}}-\frac{\partial ^{2}(\delta p+q)}{\partial y^{2}}=\nonumber\\
&& -2\eta \frac{\partial ^{2} p_{0}}{\partial x^{2}}+\frac{\partial ^{2} p_{0}}{\partial y^{2}}.
\label{eqn:eq7}
\end{eqnarray}
In this case, the right-hand side source functions are not given by the original source function, which is often a point source. On the contrary, it is related to the model perturbation and the background wavefield, which act as a secondary source possibly expanding the full space domain. 

\subsection{The physics-informed neural networks}

We use a fully connected deep neural network (NN), and the input data to the NN are spatial coordinate values. The target output parameters of the network are the real and imaginary parts of the complex scattered wavefield $\delta p$ and auxiliary wavefield $q$ corresponding to equation \ref{eqn:eq6}. The activation function between layers in the neural network is an inverse tangent function, whereas the last layer is linear. With the help of the concept of automatic differentiation \cite{baydin2017automatic}, the second-order derivatives with respect to spatial coordinates can be easily calculated. Thus, to train the network, we use the following loss function:
\begin{eqnarray}
&& f=\frac{1}{N}\sum_{i=1}^{N}|\omega ^{2}m_{n}^{(i)}\delta p^{(i)}+\frac{\partial ^{2}(\delta p^{(i))}+q^{(i)})}{\partial x^{2}} +  \nonumber\\
&&\frac{\partial ^{2}(\delta p^{(i)}+q^{(i)})}{\partial y^{2}} + \frac{1}{1+2\delta^{(i)} }\frac{\partial ^{2}\delta p^{(i)}}{\partial z^{2}} + \omega ^{2}\delta m_{n}^{(i)} p_{0}^{(i)}+ \nonumber\\
&& (\frac{1}{1+2\delta^{(i)} }-1)\frac{\partial ^{2}p_{0}^{(i)}}{\partial z^{2}}+\omega ^{2}m_{n0}^{(i)}q^{(i)}+\nonumber\\
&& 2\eta^{(i)} \frac{\partial ^{2}(\delta p^{(i)}+q^{(i)}+p_{0}^{(i)})}{\partial x^{2}}-\nonumber\\
&&\frac{\partial ^{2}(\delta p^{(i)}+q^{(i)}+p_{0}^{(i)})}{\partial y^{2}}|_{2}^{2},
\label{eqn:eq8}
\end{eqnarray}
which aims at minimizing the physics-constrained mean squared error. $N$ is the number of training points. The background wavefield $p_{0}$, true ($m_{n}$) and background ($m_{n0}$) NMO squared slowness, and anisotropic parameters $\eta$ and $\delta$ are complementary variables used in the loss function. We chose to optimize the loss function using an Adam optimizer and a follow-up L-BFGS optimization, a quasi-Newton approach, full-batch gradient-based optimization algorithm \cite{liu1989limited}.

\section{Numerical Tests}

We now share the results from implementing this approach on an anomaly model and a layered model. We compare PINN predicted results with the numerical wavefield solutions from a 9-point finite-difference wave equation operator. Finally, we test the proposed method on a modified 3D Overthrust model. The source function we use for all the examples in this paper is a delta function.

\subsection{A VTI anomaly model}

\begin{figure}
\begin{center}
\includegraphics[width=1.0\textwidth]{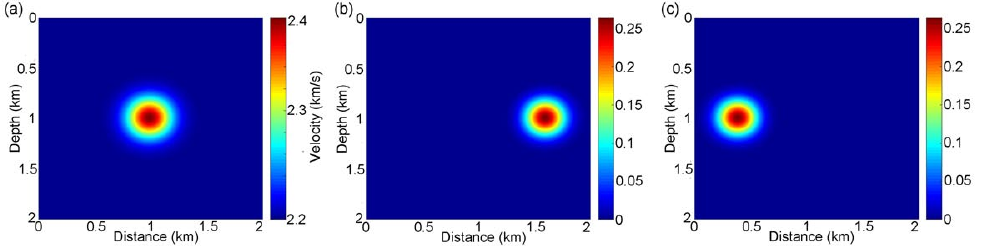} 
\caption{The true velocity (a), $\delta$ (b), and $\eta$ (c) for the VTI anomalies model.}
\label{fig:anomaly_true}
\end{center}   
\end{figure}

We first test the proposed method on parameter anomalies in the model. We show the true velocity, $\delta$, and $\eta$ models in Figs.~\ref{fig:anomaly_true}a-\ref{fig:anomaly_true}c. The anomalies are located at different locations laterally in a homogeneous isotropic background model. Thus, the background of $\delta$ and $\eta$ models is zero. The size of the models is $100 \times 100$ with a grid interval of 20 m in both vertical and horizontal directions. 

\begin{figure}
\begin{center}
\includegraphics[width=0.8\textwidth]{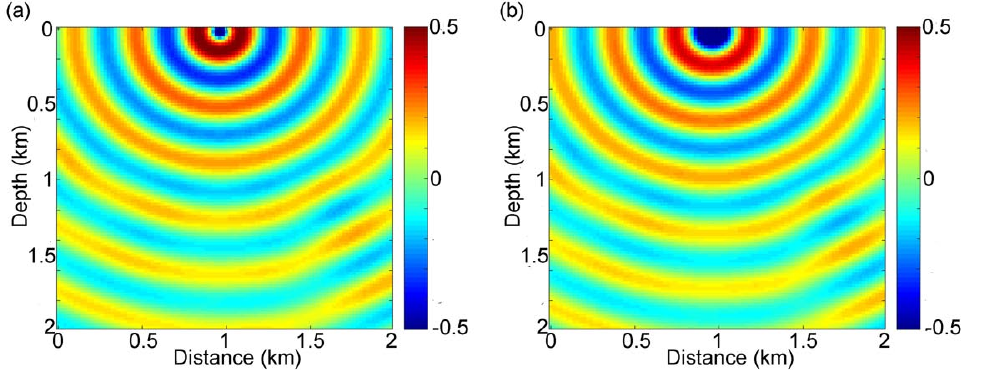} 
\caption{The real (a), and imaginary (b) parts of a 6 Hz true wavefield corresponding to the anomalies model in Figs.~\ref{fig:anomaly_true}a-\ref{fig:anomaly_true}c evaluated using a numerical method.}
\label{fig:anomaly_u}
\end{center}   
\end{figure}

\begin{figure}
\begin{center}
\includegraphics[width=0.8\textwidth]{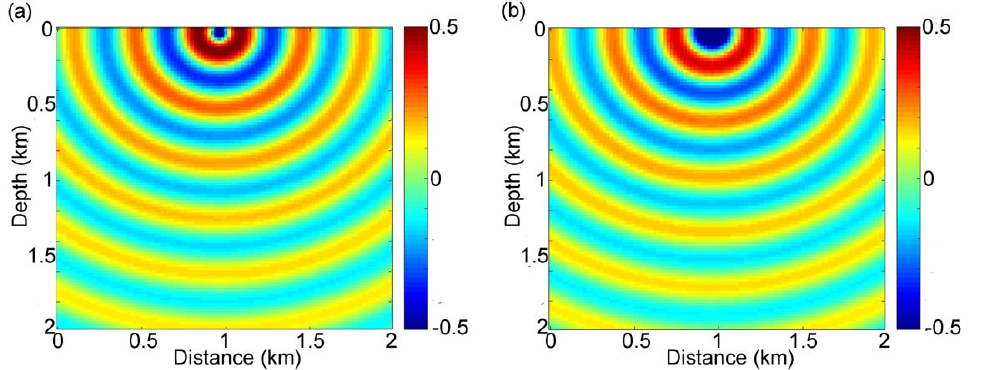} 
\caption{The real (a), and imaginary (b) parts of a 6 Hz background wavefield corresponding to a homogeneous isotropic background model evaluated analytically.}
\label{fig:anomaly_u0}
\end{center}   
\end{figure}

\begin{figure}
\begin{center}
\includegraphics[width=0.8\textwidth]{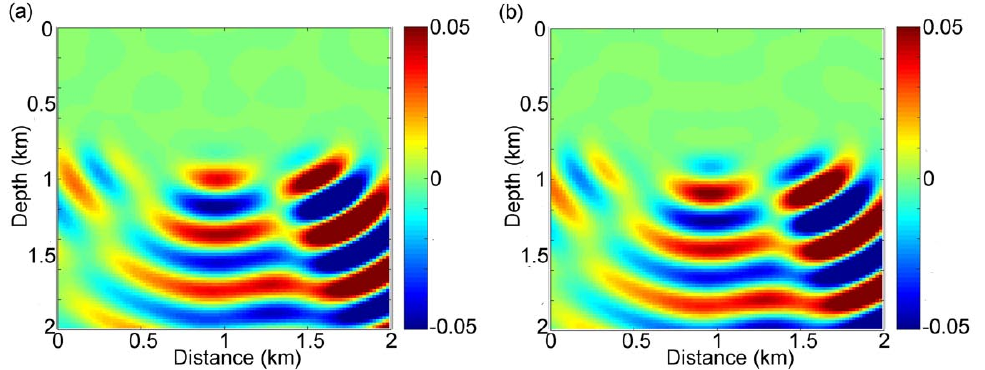} 
\caption{The real (a), and imaginary (b) parts of the perturbed (scattered) wavefield given by the difference between the wavefields in Figs.~\ref{fig:anomaly_u} and~\ref{fig:anomaly_u0}.}
\label{fig:anomaly_du}
\end{center}   
\end{figure}

We consider a point source on the surface of the model located at 1 km along the horizontal axis. The real and imaginary parts of the pressure wavefield at 6 Hz using a finite-difference method are shown in Figs.~\ref{fig:anomaly_u}a and~\ref{fig:anomaly_u}b, respectively. Using the same frequency and source function, the real and imaginary parts of the background wavefields solved analytically corresponding to the background velocity are shown in Figs.~\ref{fig:anomaly_u0}a and~\ref{fig:anomaly_u0}b, respectively. We subtract the two wavefields in Fig.~\ref{fig:anomaly_u} and Fig.~\ref{fig:anomaly_u0} to obtain the reference scattered wavefield, as shown in Figs.~\ref{fig:anomaly_du}a and~\ref{fig:anomaly_du}b, for the real and imaginary parts, respectively. As the anomalies are smooth, most of the perturbed (scattered) wavefield correspond to transmission. For transmission, considering this parameter set to represent the VTI acousitc model, \cite{alkhalifah2016research} showed that $\eta$ perturbations admits practically no transmission energy in the vertical direction, while the other parameters do. The strong $\delta$ perturbation based energy is caused from the angular deviated location with respect to the source and the relative size of the perturbation, compared to the velocity (effectively $>20\%$ compared to $10\%$). So the numerical results are consistent with the analytical expectations, so let us observe whether PINNs can be trained to do so. 

\begin{figure}
\begin{center}
\includegraphics[width=0.8\textwidth]{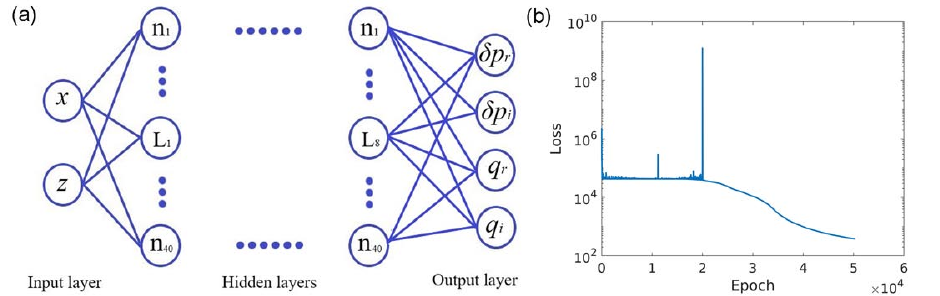} 
\caption{The PINN architecture with the input data being spatial coordinates and outputs being the real and imaginary parts of the scattered wavefield $\delta p$ and the auxiliary function $q$ (a), and the corresponding loss function curve for the training of the NN using Adam and LBFGS optimizers (b).}
\label{fig:PIDNN_misfit}
\end{center}   
\end{figure}

\begin{figure}
\begin{center}
\includegraphics[width=0.8\textwidth]{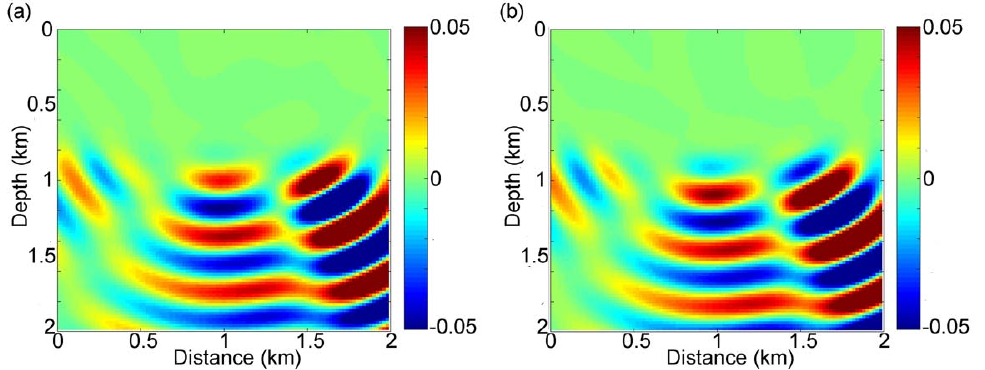} 
\caption{The real (a), and imaginary (b) parts of a 6 Hz scattered wavefield corresponding to the anomalies model using PINN.}
\label{fig:anomaly_du_ml}
\end{center}   
\end{figure}

\begin{figure}
\begin{center}
\includegraphics[width=0.8\textwidth]{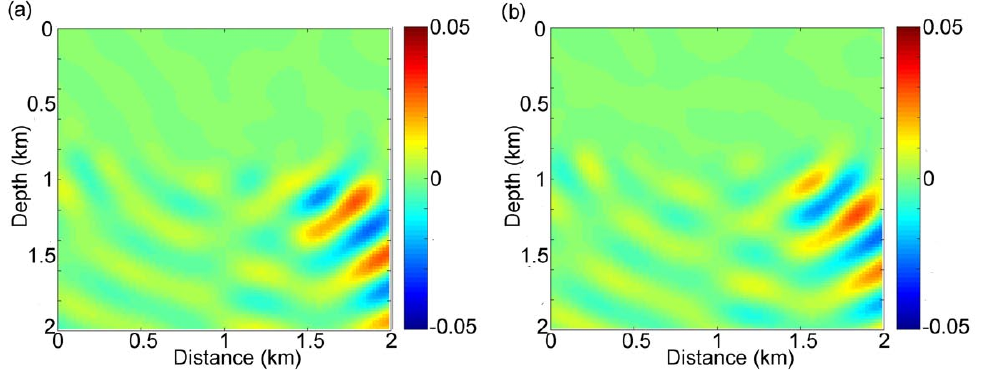} 
\caption{The real (a), and imaginary (b) parts of scattered wavefield differences between the numerical solutions and PINN predicted solutions.}
\label{fig:anomaly_du_dif}
\end{center}   
\end{figure}

For this example, we use a fully connected 8-layer deep neural network, and each layer has 40 neurons, as shown in Fig.~\ref{fig:PIDNN_misfit}a. Half of the regular grid points are randomly fed to the network. After 20,000 epochs of Adam optimizer training and 30,000 LBFGS updates, we show the loss function curve in Fig.~\ref{fig:PIDNN_misfit}b. By inputting all the regular grid points into the trained network, the real and imaginary parts of the predicted scattered wavefields are shown in Figs.~\ref{fig:anomaly_du_ml}a and~\ref{fig:anomaly_du_ml}b, respectively. It is difficult to distinguish the difference between the numerical method and PINNs predicted solutions. We subtract the two solutions, and show the difference plotted at the same scale in Fig.~\ref{fig:anomaly_du_dif}. It is obvious that the perturbations in the wavefiels due to velocity and $\eta$ have been accurately simulated, and only mild differences are observed focused on the $\delta$ perturbation due to its strength. This difference is mainly effecting the amplitude slightly only and not in the phase.

\subsection{A layered VTI model}

\begin{figure}
\begin{center}
\includegraphics[width=0.8\textwidth]{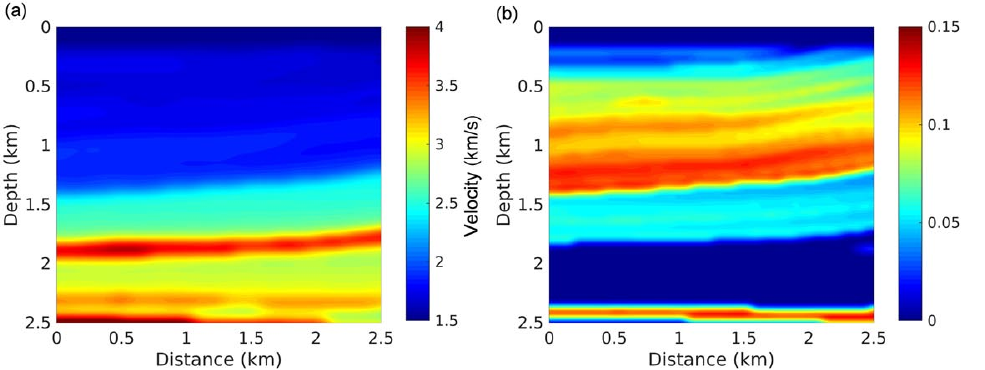} 
\caption{The true velocity (a), $\delta$ and $\eta$ (b) of a layered models.} 
\label{fig:mar_true}
\end{center}   
\end{figure}

Next, we consider a simple layered model, which is extracted from the left side of the anisotropic Marmousi model and slightly smoothed. We set a shallow water layer on the top of the model. The true velocity is shown in Fig.~\ref{fig:mar_true}a. We use the same model for $\delta$ and $\eta$, given in Fig.~\ref{fig:mar_true}b. The model has 100 samples in both the horizontal and vertical directions with a grid interval of 25 m.

\begin{figure}
\begin{center}
\includegraphics[width=1.0\textwidth]{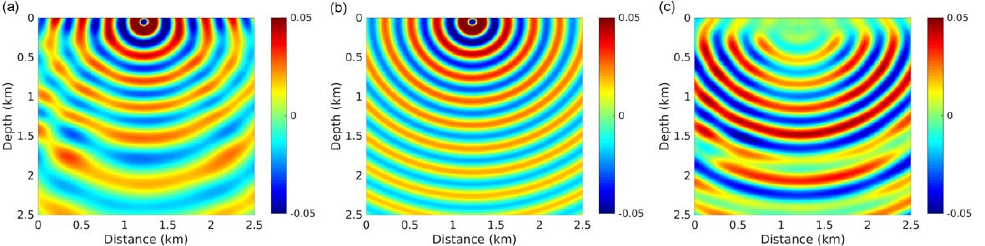} 
\caption{The real part of a 5 Hz true wavefield using a numerical method (a), background wavefield solved analytically (b), and the scattered wavefield (c) given by difference between the two wavefields.} 
\label{fig:mar_u_u0_du}
\end{center}   
\end{figure}

We again place the source on the surface at location 1.25 km, which corresponds to an the isotropic water layer. The real part of the numerical wavefield solution for 5 Hz is shown in Fig.~\ref{fig:mar_u_u0_du}a. We solve the background wavefield analytically with a homogeneous velocity of 1.5 km/s and show the real part in Fig.~\ref{fig:mar_u_u0_du}b. The difference between the numerical and background wavefields is considered as the reference solution for the scattered wavefield, which is shown in Fig.~\ref{fig:mar_u_u0_du}c. 

\begin{figure}
\begin{center}
\includegraphics[width=1.0\textwidth]{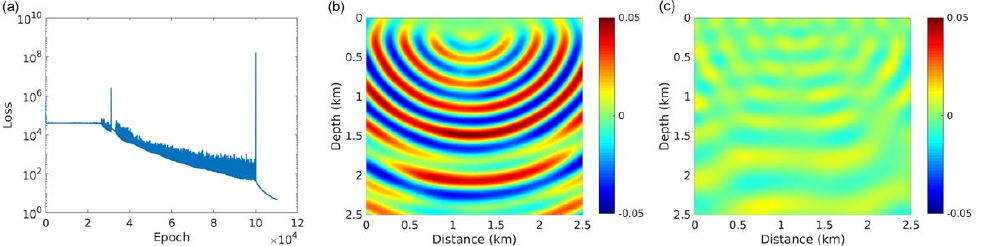} 
\caption{The loss function curve for the training of the NN using Adam and LBFGS optimizers (a), the scattered wavefield using PINN method, and the scattered wavefield difference between Fig.~\ref{fig:mar_u_u0_du}c and Fig.~\ref{fig:mar_mis_du_dif}b (c).} 
\label{fig:mar_mis_du_dif}
\end{center}   
\end{figure}

We use the same PINN architecture as in the previous example. In this case, we only use 2000 random training points to train the network. We use 100,000 epochs of Adam optimizer and 15,000 iterations of LBFGS to update the parameters in the network and show the loss function curve in Fig.~\ref{fig:mar_mis_du_dif}a. By inputting the spatial coordinate values of the regular grid points, used in the numerical solutions, to the network, the real part of the predicted scattered wavefield is shown in Fig.~\ref{fig:mar_mis_du_dif}b. We observe that the general shape of the scattered wavefield from numerical and the PINNs predicted solutions are very close. Again, we show the difference between the numerical and PINNs predicted scattered wavefields in Fig.~\ref{fig:mar_mis_du_dif}c. Note that in most of the areas, the scattered wavefield differences between the two methods are quite small.  

\begin{figure}
\begin{center}
\includegraphics[width=1.0\textwidth]{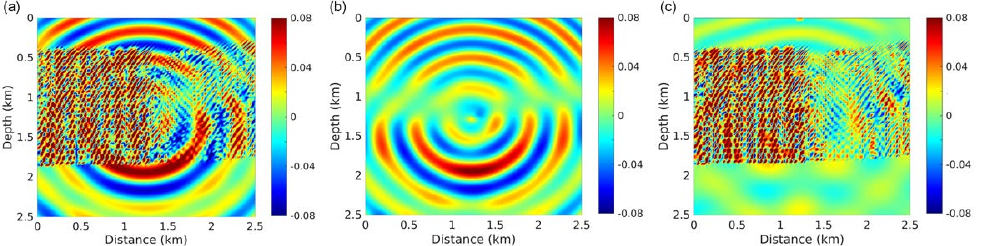} 
\caption{The real part of the scattered wavefield using a numerical method (a), the scattered wavefield using PINN (b), and their difference (c) when the source is in the center.}  
\label{fig:mar_du_duml_dif}
\end{center}   
\end{figure}

Next, we place the source in the center of the model, where strong anisotropy exists. It was shown that the wavefield solution from acoustic anisotropic wave equation suffers from shear wave artifacts using a finite-difference solver, and this issue will be more severe when the source is located in the anisotropic region \cite{alkhalifah2000acoustic,wu2016waveform}. In  this case, the background wavefield solved analytically uses a homogeneous velocity of 1.9 km/s. The real part of the resulting scattered wavefield is shown in Fig.~\ref{fig:mar_du_duml_dif}a. It is obvious that there are strong high-wavenumber artifacts in the low-velocity areas. This effect can be explained by the dispersion relation $\omega=v(\mathbf{x},k)$ with $k$ denoting the wavenumber. As shear wave artifacts are given by waves with relatively low velocity, the artifacts tend to be given by high wavenumber components \cite{wu2016waveform}. Using the same training setup in this model, we train the network and output the predicted wavefield, as shown in Fig.~\ref{fig:mar_du_duml_dif}b. We observe that the PINNs predicted wavefield is free of the shear wave artifacts. The difference between scattered wavefields using numerical and PINN methods is shown Fig.~\ref{fig:mar_du_duml_dif}c. We observe that the difference in the background is small, mainly due to the amplitude difference, and the high-wavenumber artifacts are isolated from the numerical solutions.

\begin{figure}
\begin{center}
\includegraphics[width=0.8\textwidth]{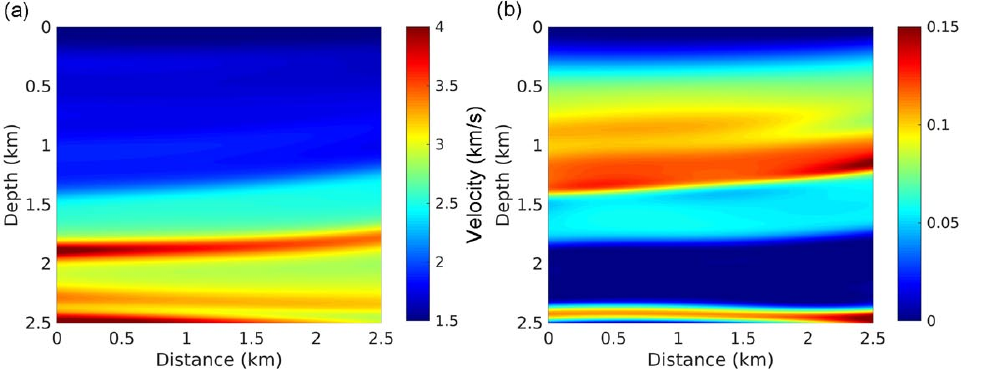} 
\caption{The predicted velocity (a), and $\delta$ (b) corresponding to the PINN predicted scattered wavefield.}  
\label{fig:mar_v_delta_pred}
\end{center}   
\end{figure}

To further verify the accuracy of the PINNs predicted wavefield, we predict the velocity and $\delta$ using the predicted scattered wavefields using another independent network \cite{song2020pinnwri}. In this new network, we still use spatial coordinates as the input data. We use the first equation in equation \ref{eqn:eq7} as the loss function. As the scattered and perturbation wavefields are known parameters, the target output parameters are NMO squared slowness and $\delta$. This new network is rather small, which has only five hidden layers and 20 neurons in each layer. Only 1000 epochs of Adam optimizer is used to train the network. The results are shown in Fig.~\ref{fig:mar_v_delta_pred}a and~\ref{fig:mar_v_delta_pred}b, respectively. We observe that the predicted velocity and $\delta$ models using the predicted wavefields are close to the true models, only slightly smoothed, which demonstrates the accuracy of the predicted scattered wavefields.

\subsection{A modified Overthrust model}

\begin{figure}
\begin{center}
\includegraphics[width=1.0\textwidth]{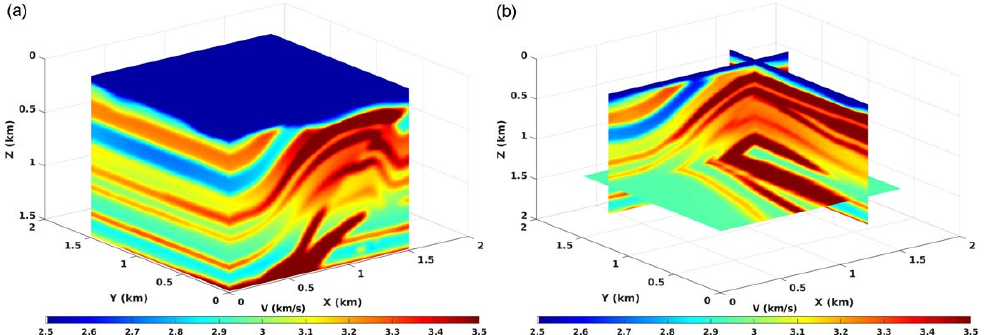} 
\caption{The velocity model (a) and three corresponding slices (b) of the modified Overthrust model.}  
\label{fig:over3d_v}
\end{center}   
\end{figure}

\begin{figure}
\begin{center}
\includegraphics[width=1.0\textwidth]{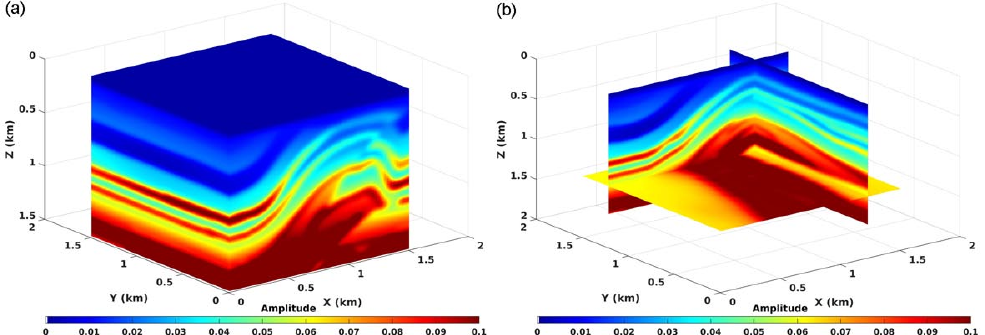} 
\caption{The $\eta$  model (a) and three corresponding slices (b) of the modified Overthrust model.}  
\label{fig:over3d_eta}
\end{center}   
\end{figure}

\begin{figure}
\begin{center}
\includegraphics[width=1.0\textwidth]{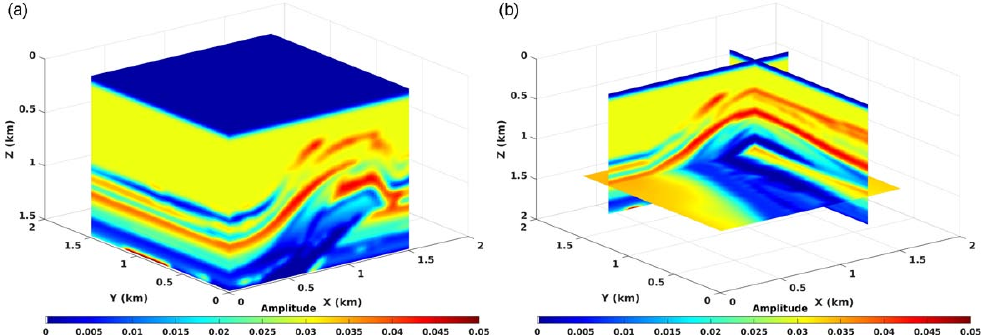} 
\caption{The $\delta$  model (a) and three corresponding slices (b) of the modified Overthrust model.}  
\label{fig:over3d_delta}
\end{center}   
\end{figure}

\begin{figure}
\begin{center}
\includegraphics[width=1.0\textwidth]{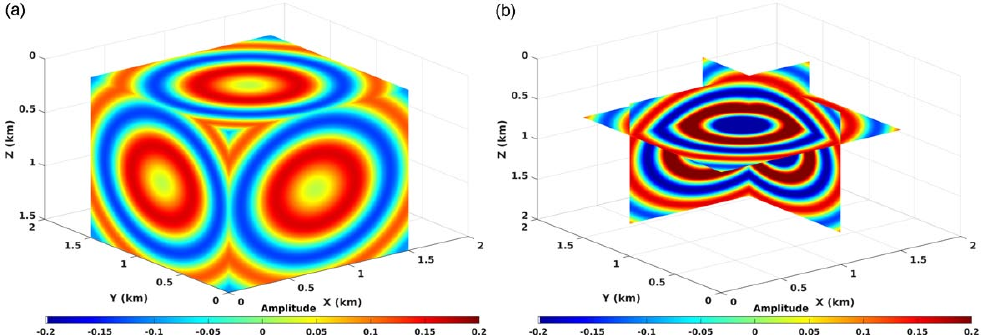} 
\caption{The real part of the 10 Hz background wavefield (a), and three slices corresponding to it (b).}
\label{fig:over3d_u0}
\end{center}   
\end{figure}

We further test the proposed method on a 3D modified anisotropic Overthrust model. The velocity model is shown in Fig.~\ref{fig:over3d_v}a. To highlight the details of the model, three inner slices are shown in Fig.~\ref{fig:over3d_v}b. We show the $\eta$ and $\delta$ models with the same plot configuration in Figs.~\ref{fig:over3d_eta} and~\ref{fig:over3d_delta}, respectively. This model has 61 samples in the $x$, $y$, and $z$ directions with a 25 m sampling interval. The source is in the center of the model, and the resulting real part of the 10 Hz background wavefield calculated analytically for a homogeneous velocity of 3.2 km/s is shown in Fig.~\ref{fig:over3d_u0}.

\begin{figure}
\begin{center}
\includegraphics[width=1.0\textwidth]{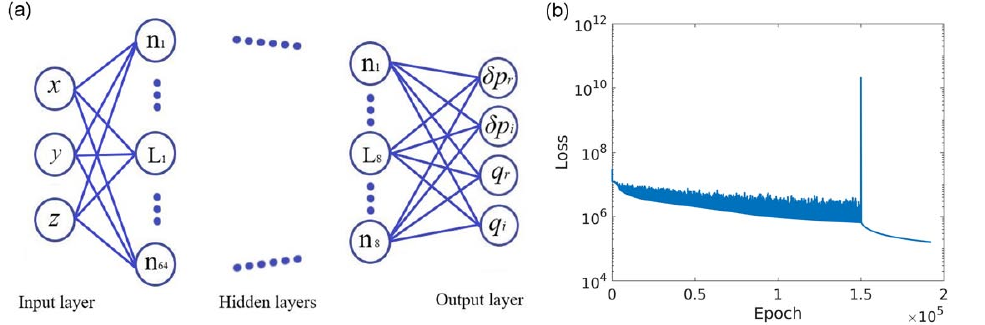} 
\caption{The PINN architecture (a), and the corresponding loss function curve for the training of the NN using Adam and LBFGS optimizers (b) used for the modified Overthrust model.}
\label{fig:PIDNN_misfit_over3d}
\end{center}   
\end{figure}

\begin{figure}
\begin{center}
\includegraphics[width=1.0\textwidth]{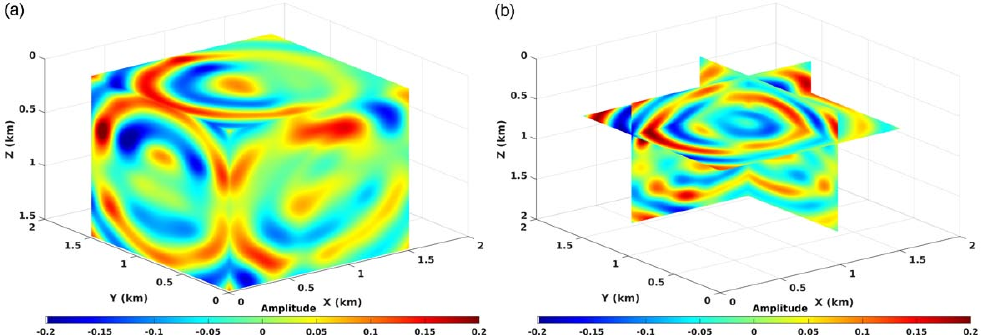} 
\caption{The real part of the scattered wavefield using PINN (a), and three slices corresponding to it (b).}
\label{fig:over3d_du_ml}
\end{center}   
\end{figure}

\begin{figure}
\begin{center}
\includegraphics[width=1.0\textwidth]{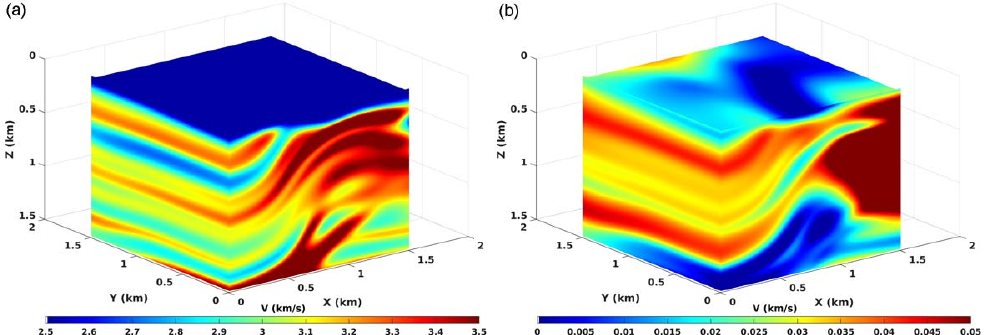} 
\caption{The predicted velocity (a), and $\delta$ (b) corresponding to the PINN predicted scattered wavefield in Figs.~\ref{fig:over3d_du_ml}a and~\ref{fig:over3d_du_ml}b.}
\label{fig:over3d_pred_v_delta}
\end{center}   
\end{figure}

The network used in this example has eight layers with $\left \{ 64, 64, 32, 32, 16, 16, 8, 8 \right \}$ neurons from shallow to deep, as shown in Fig.~\ref{fig:PIDNN_misfit_over3d}a. We choose 50,000 randomly sampled points from the regular grid in the training process. Adam optimizer with 150,000 epochs and a follow-up LBFGS with 40,000 updates are used to train this network. The corresponding loss function curve is shown in Fig.~\ref{fig:PIDNN_misfit_over3d}b. The real part of the predicted scattered wavefield is shown in Fig.~\ref{fig:over3d_du_ml}. We predict the velocity and $\delta$ models using the predicted scattered wavefields with the same model prediction network as the last example. The resulting velocity and $\delta$ models are shown in Figs.~\ref{fig:over3d_pred_v_delta}a and~\ref{fig:over3d_pred_v_delta}b, respectively. They are close to the true models, which confirms the accuracy of the predicted scattered wavefields.

\subsection{Topography model}

\begin{figure}
\begin{center}
\includegraphics[width=0.8\textwidth]{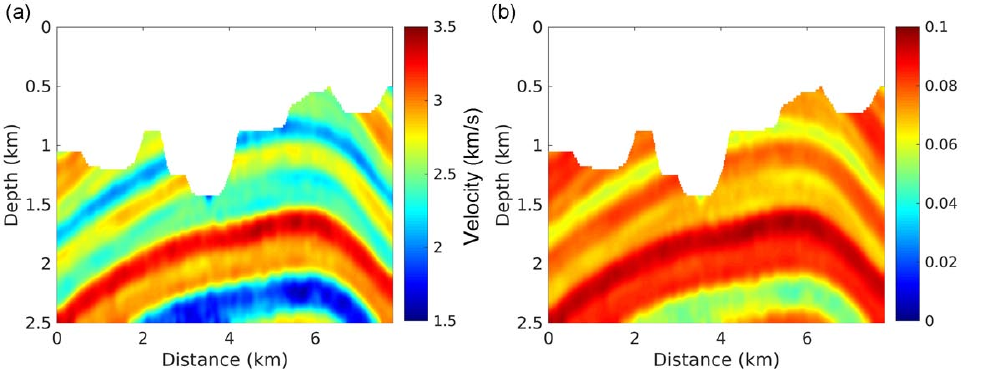} 
\caption{The true velocity (a), $\delta$ and $\eta$ (b) of a model with irregular topography.}
\label{fig:topo_v_eta}
\end{center}   
\end{figure}

\begin{figure}
\begin{center}
\includegraphics[width=0.8\textwidth]{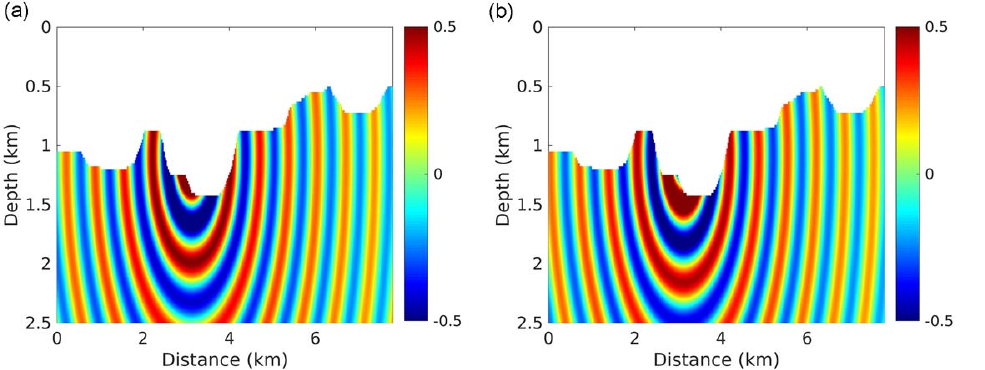} 
\caption{The real (a) and imaginary (b) parts of the 3 Hz background wavefield.}
\label{fig:topo_u0}
\end{center}   
\end{figure}

Finally, we apply the proposed method on a velocity model with irregular topography, which is extracted from the SEAM foothills model \cite{regone2017geologic}. We build the anisotropic models, $\eta$ and $\delta$, by dividing the velocity model by a constant value. The velocity and anisotropic models are shown in Figs.~\ref{fig:topo_v_eta}a and~\ref{fig:topo_v_eta}b, respectively. We consider a source located at (1.28 km, 3.13 km) and use 2.0 km/s as the homogeneous isotropic background velocity. The corresponding real and imaginary parts of the 3 Hz background wavefield are shown in Figs.~\ref{fig:topo_u0}a and~\ref{fig:topo_u0}b, respectively.

\begin{figure}
\begin{center}
\includegraphics[width=0.8\textwidth]{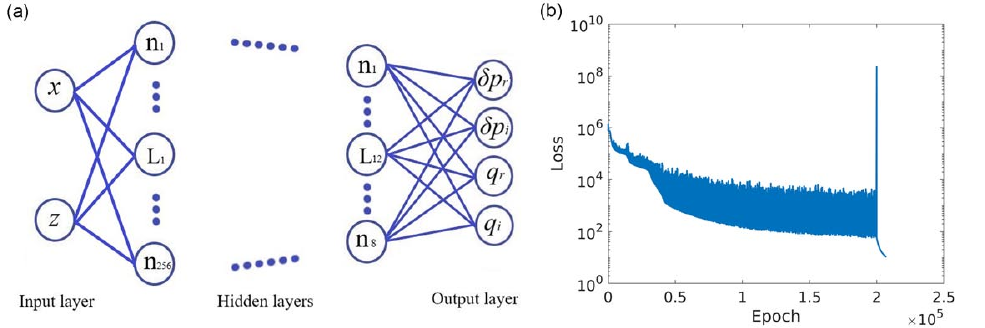} 
\caption{The PINN architecture (a), and the corresponding loss function curve for the training of the NN using Adam and LBFGS optimizers (b) used for the model with irregular topography.}
\label{fig:topo_pinn_mis}
\end{center}   
\end{figure}

\begin{figure}
\begin{center}
\includegraphics[width=0.8\textwidth]{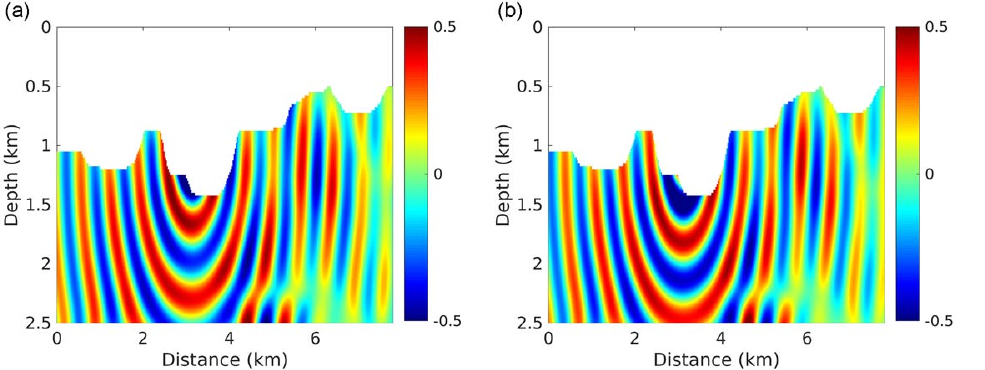} 
\caption{The real (a) and imaginary (b) parts of the 3 Hz predicted scattered wavefield using PINN.}
\label{fig:topo_du}
\end{center}   
\end{figure}

\begin{figure}
\begin{center}
\includegraphics[width=0.8\textwidth]{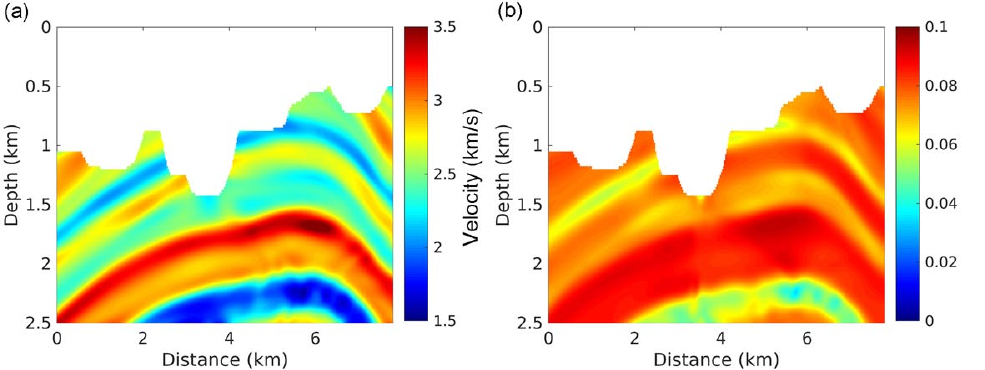} 
\caption{The predicted velocity (a), and $\delta$ (b) corresponding to the PINN predicted scattered wavefield in Figs.~\ref{fig:topo_du}a and~\ref{fig:topo_du}b.}
\label{fig:topo_v_delta_pred}
\end{center}   
\end{figure}

We feed 10,000 random points to a 12-layer network with $\left \{256, 256, 128, 128, 64, 64, 32, 32, 16, 16, 8, 8 \right \}$ neurons from shallow to deep, as shown in Fig.~\ref{fig:topo_pinn_mis}a. We train this network using the Adam optimizer with 200,000 epochs and a follow-up LBFGS with 10,000 updates and show the corresponding loss function curve in Fig.~\ref{fig:topo_pinn_mis}b. After the network is trained, it acts as a function to produce the real and imaginary parts of the scattered wavefield, which are shown in Figs.~\ref{fig:topo_du}a and~\ref{fig:topo_du}b, respectively. Again, we use the same model prediction network to predict the velocity and $\delta$. The results are shown in Figs.~\ref{fig:topo_v_delta_pred}a and~\ref{fig:topo_v_delta_pred}b, respectively. Compared to the true models in Fig.~\ref{fig:topo_v_eta}, the accurate predicted models demonstrate the accuracy of the predicted scattered wavefields. 

\section{Discussion}

Instead of using the acoustic VTI wave equation directly as the loss function, we use a variation of it to solve for the scattered pressure wavefield. The source function used in the frequency-domain wave equation modeling is highly sparse, and non-zero values only exist at the source location. This sparse feature of the source function will require more training samples at the source area to properly capture the large curvature of the wavefield there, which will hamper the training convergence.We use a background homogeneous and isotropic velocity to analytically obtain the background wavefield. We recommend using a velocity given by the source location velocity so the background wavefield would absorb the majority of the source singularity effect. If the source is located in the water layer, like in a marine survey, we can just use the water velocity to generate the background wavefields. In this case, the target scattered wavefields do not have specially large values near the source location.

With the increase in the velocity complexity, we need to use a bigger network or a larger number of epochs to train the network. Otherwise, we will end up with smoother wavefield solutions. However, smoother wavefields may be beneficial for velocity inversion like FWI, as FWI favors smooth gradient updates in the early stages \cite{vir}. The loss function used to train the network can easily include the data fitting term, and this feature provides the potential for solving an augmented wave equation and performing a wavefield reconstruction inversion \cite{van2013mitigating}. As we aim at solving an acoustic VTI wave equation, we may provide an alternative way of implementing a multiparameter inversion for acoustic VTI media.

We use the predicted scattered wavefields to reconstruct the velocity and $\delta$ models by building another neural network when the numerical solutions of the predicted scattered wavefields are difficult to obtain. This new network is small and easy to train, as all the wavefields and their derivatives are calculated before training. We observe that the predicted models are slightly smoothed. This is because the limitations of the network size and training points number result in smoother predicted wavefields than the true ones.

\section{Conclusion}

We developed a novel approach to solve the frequency-domain acoustic VTI wave equation using physics-informed neural networks (PINNs). We design a fully connected neural network, which accepts spatial coordinates as input data. The target output parameters of the network are the real and imaginary parts of the scattered pressure and auxiliary wavefields. By using PINNs to obtain the scattered wavefield solutions, we avoid calculating the inverse of the impedance matrix, which is computationally expensive for large models. The proposed method is also immune to the shear wave artifacts. The trained network acts as a function of spatial coordinate values, so it allows for models in any shapes. Applications on 2D and 3D models show that the proposed method can generate the scattered wavefield solutions with reasonable accuracy.

\section{Acknowledgement}

We thank KAUST for its support and the SWAG group for the collaborative environment. This work utilized the resources of the Supercomputing Laboratory at King Abdullah University of Science and Technology (KAUST) in Thuwal, Saudi Arabia, and we are grateful for that. We thank Bin She, from the University of electronic science and technology, for sharing the 3D plot tool. 

\bibliographystyle{unsrt}  
\bibliography{refs/pinn_helm_vti}

\begin{thebibliography}{10}

\bibitem{alkhalifah2000acoustic}
Tariq Alkhalifah.
\newblock An acoustic wave equation for anisotropic media.
\newblock {\em Geophysics}, 65(4):1239--1250, 2000.

\bibitem{alkhalifah1998acoustic}
Tariq Alkhalifah.
\newblock Acoustic approximations for processing in transversely isotropic
  media.
\newblock {\em Geophysics}, 63(2):623--631, 1998.

\bibitem{zhou2006anisotropic}
H~Zhou, G~Zhang, and R~Bloor.
\newblock An anisotropic acoustic wave equation for vti media.
\newblock In {\em 68th EAGE Conference and Exhibition incorporating SPE EUROPEC
  2006}, pages cp--2. European Association of Geoscientists \& Engineers, 2006.

\bibitem{song2020efficient}
Chao Song and Tariq Alkhalifah.
\newblock An efficient wavefield inversion for transversely isotropic media
  with a vertical axis of symmetry.
\newblock {\em Geophysics}, 85(3):R195--R206, 2020.

\bibitem{shi2018microseismic}
Peidong Shi, Doug Angus, Andy Nowacki, Sanyi Yuan, and Yanyan Wang.
\newblock Microseismic full waveform modeling in anisotropic media with moment
  tensor implementation.
\newblock {\em Surveys in Geophysics}, 39(4):567--611, 2018.

\bibitem{wu2016waveform}
Zedong Wu and Tariq Alkhalifah.
\newblock Waveform inversion for acoustic vti media in frequency domain.
\newblock In {\em SEG Technical Program Expanded Abstracts 2016}, pages
  1184--1189. Society of Exploration Geophysicists, 2016.

\bibitem{vapnik2013nature}
Vladimir Vapnik.
\newblock {\em The nature of statistical learning theory}.
\newblock Springer science \& business media, 2013.

\bibitem{li2004support}
Jiakang Li and John Castagna.
\newblock Support vector machine (svm) pattern recognition to avo
  classification.
\newblock {\em Geophysical Research Letters}, 31(2), 2004.

\bibitem{zhao2015comparison}
Tao Zhao, Vikram Jayaram, Atish Roy, and Kurt~J Marfurt.
\newblock A comparison of classification techniques for seismic facies
  recognition.
\newblock {\em Interpretation}, 3(4):SAE29--SAE58, 2015.

\bibitem{song2018source}
Chao Song, T~Alkhalifah, and Z~Wu.
\newblock Source type classification based on the support vector machine
  method.
\newblock In {\em 80th EAGE Conference and Exhibition 2018}, volume 2018, pages
  1--5. European Association of Geoscientists \& Engineers, 2018.

\bibitem{microsvm}
C~Song and T~Alkhalifah.
\newblock Identifying microseismic events in time-reversed source images using
  support vector machine.
\newblock In {\em 82nd EAGE Annual Conference \& Exhibition}, volume 2020,
  pages 1--5. European Association of Geoscientists \& Engineers, 2020.

\bibitem{chen2020suppressing}
Yuqing Chen, Yunsong Huang, and Lianjie Huang.
\newblock Suppressing migration image artifacts using a support vector machine
  method.
\newblock {\em Geophysics}, 85(5):1--55, 2020.

\bibitem{shi2018automatic}
Yunzhi Shi, Xinming Wu, and Sergey Fomel.
\newblock Automatic salt-body classification using a deep convolutional neural
  network.
\newblock In {\em SEG Technical Program Expanded Abstracts 2018}, pages
  1971--1975. Society of Exploration Geophysicists, 2018.

\bibitem{wu2019faultseg3d}
Xinming Wu, Luming Liang, Yunzhi Shi, and Sergey Fomel.
\newblock Faultseg3d: Using synthetic data sets to train an end-to-end
  convolutional neural network for 3d seismic fault segmentation.
\newblock {\em Geophysics}, 84(3):IM35--IM45, 2019.

\bibitem{ovcharenko2019deep}
Oleg Ovcharenko, Vladimir Kazei, Mahesh Kalita, Daniel Peter, and Tariq
  Alkhalifah.
\newblock Deep learning for low-frequency extrapolation from multioffset
  seismic data.
\newblock {\em Geophysics}, 84(6):R989--R1001, 2019.

\bibitem{regone2017geologic}
Carl Regone, Joseph Stefani, Peter Wang, Constantin Gerea, Gladys Gonzalez, and
  Michael Oristaglio.
\newblock Geologic model building in seam phase ii—land seismic challenges.
\newblock {\em The Leading Edge}, 36(9):738--749, 2017.

\bibitem{zhu2019phasenet}
Weiqiang Zhu and Gregory~C Beroza.
\newblock Phasenet: a deep-neural-network-based seismic arrival-time picking
  method.
\newblock {\em Geophysical Journal International}, 216(1):261--273, 2019.

\bibitem{kaur2020improving}
Harpreet Kaur, Nam Pham, and Sergey Fomel.
\newblock Improving resolution of migrated images by approximating the inverse
  hessian using deep learning.
\newblock {\em Geophysics}, 85(4):1--62, 2020.

\bibitem{zhang2020high}
Zhen-dong Zhang and Tariq Alkhalifah.
\newblock High-resolution reservoir characterization using deep learning-aided
  elastic full-waveform inversion: The north sea field data example.
\newblock {\em Geophysics}, 85(4):WA137--WA146, 2020.

\bibitem{liyy}
Y~Li, T~Alkhalifah, and Z~Zhang.
\newblock High-resolution regularized elastic full waveform inversion assisted
  by deep learning.
\newblock In {\em 82nd EAGE Annual Conference \& Exhibition}, volume 2020,
  pages 1--5. European Association of Geoscientists \& Engineers, 2020.

\bibitem{siahkoohi2019TRnna}
Ali Siahkoohi, Mathias Louboutin, and Felix~J. Herrmann.
\newblock Neural network augmented wave-equation simulation.
\newblock 09 2019.

\bibitem{moseley2020solving}
Ben Moseley, Andrew Markham, and Tarje Nissen-Meyer.
\newblock Solving the wave equation with physics-informed deep learning.
\newblock {\em arXiv preprint arXiv:2006.11894}, 2020.

\bibitem{8931232}
S.~{Li}, B.~{Liu}, Y.~{Ren}, Y.~{Chen}, S.~{Yang}, Y.~{Wang}, and P.~{Jiang}.
\newblock Deep-learning inversion of seismic data.
\newblock {\em IEEE Transactions on Geoscience and Remote Sensing},
  58(3):2135--2149, 2020.

\bibitem{ren2020physics}
Yuxiao Ren, Xinji Xu, Senlin Yang, Lichao Nie, and Yangkang Chen.
\newblock A physics-based neural-network way to perform seismic full waveform
  inversion.
\newblock {\em IEEE Access}, 8:112266--112277, 2020.

\bibitem{raissi2019physics}
Maziar Raissi, Paris Perdikaris, and George~E Karniadakis.
\newblock Physics-informed neural networks: A deep learning framework for
  solving forward and inverse problems involving nonlinear partial differential
  equations.
\newblock {\em Journal of Computational Physics}, 378:686--707, 2019.

\bibitem{sahli2020physics}
Francisco Sahli~Costabal, Yibo Yang, Paris Perdikaris, Daniel~E Hurtado, and
  Ellen Kuhl.
\newblock Physics-informed neural networks for cardiac activation mapping.
\newblock {\em Frontiers in Physics}, 8:42, 2020.

\bibitem{raissi2020hidden}
Maziar Raissi, Alireza Yazdani, and George~Em Karniadakis.
\newblock Hidden fluid mechanics: Learning velocity and pressure fields from
  flow visualizations.
\newblock {\em Science}, 367(6481):1026--1030, 2020.

\bibitem{helmholtz}
T~Alkhalifah, C~Song, Q~Hao, et~al.
\newblock Wavefield solutions from machine learned functions that approximately
  satisfy the wave equation.
\newblock In {\em 82nd EAGE Annual Conference \& Exhibition}, volume 2020,
  pages 1--5. European Association of Geoscientists \& Engineers, 2020.

\bibitem{eikonal}
UB~Waheed, Ehsan Haghighat, Tariq Alkhalifah, Chao Song, and Qi~Hao.
\newblock Eikonal solution using physics-informed neural networks.
\newblock In {\em 82nd EAGE Annual Conference \& Exhibition}, volume 2020,
  pages 1--5. European Association of Geoscientists \& Engineers, 2020.

\bibitem{engquist2018approximate}
Bj{\"o}rn Engquist and Hongkai Zhao.
\newblock Approximate separability of the green's function of the helmholtz
  equation in the high frequency limit.
\newblock {\em Communications on Pure and Applied Mathematics},
  71(11):2220--2274, 2018.

\bibitem{baydin2017automatic}
At{\i}l{\i}m~G{\"u}nes Baydin, Barak~A Pearlmutter, Alexey~Andreyevich Radul,
  and Jeffrey~Mark Siskind.
\newblock Automatic differentiation in machine learning: a survey.
\newblock {\em The Journal of Machine Learning Research}, 18(1):5595--5637,
  2017.

\bibitem{liu1989limited}
Dong~C Liu and Jorge Nocedal.
\newblock On the limited memory bfgs method for large scale optimization.
\newblock {\em Mathematical programming}, 45(1-3):503--528, 1989.

\bibitem{alkhalifah2016research}
Tariq Alkhalifah.
\newblock Research note: Insights into the data dependency on anisotropy: An
  inversion prospective.
\newblock {\em Geophysical Prospecting}, 64(2):505--513, 2016.

\bibitem{song2020pinnwri}
Chao Song and Tariq Alkhalifah.
\newblock Wavefield reconstruction inversion via machine learned functions.
\newblock In {\em SEG Technical Program Expanded Abstracts 2020}. Society of
  Exploration Geophysicists, 2020.

\bibitem{vir}
Jean Virieux and St{\'e}phane Operto.
\newblock An overview of full-waveform inversion in exploration geophysics.
\newblock {\em Geophysics}, 74(6):WCC1--WCC26, 2009.

\bibitem{van2013mitigating}
Tristan Van~Leeuwen and Felix~J Herrmann.
\newblock Mitigating local minima in full-waveform inversion by expanding the
  search space.
\newblock {\em Geophysical Journal International}, 195(1):661--667, 2013.

\end{thebibliography}

\end{document}